\documentclass[useAMS,usenatbib,usegraphicx]{mn2e}

\usepackage{amsmath,amssymb,amsbsy,amsfonts}
\usepackage{amssymb}
\usepackage[british]{babel}
\usepackage{mdwmath}
\usepackage{color}
\usepackage{url}

\title{Internal shocks in relativistic jets with time--dependent sources}
\author[S. Mendoza, J.C. Hidalgo, D. Olvera \& J.I. Cabrera]
       {S. Mendoza\(^{1}\), J.C. Hidalgo\(^{2} \), D. Olvera\(^{1}\) \&
           J.I. Cabrera\(^{1}\) \\
               \(^1\) Instituto de Astronom\'{\i}a, Universidad Nacional
           Aut\'onoma de M\'exico, AP 70-264, Distrito Federal 04510,
           M\'exico\\
               \(^2\) Astronomy Unit, Queen Mary College, University of London,
           Mile End Road, London E1 4NS, United Kingdom 
           }

\begin{document}

\maketitle
\label{firstpage}
\begin{abstract}
  We present a ballistic description of the formation and propagation
of the working surface of a relativistic jet.  Using simple laws of
conservation of mass and linear momentum at the working surface, we
obtain a full description of the working surface flow parametrised by
the initial velocity and mass injection rate. This allows us to compute
analytically the energy release at any time in the working surface. We
compare this model with the results obtained numerically through a new
hydrodynamical code applied to the propagation of a relativistic fluid
in one dimension in order to test the limits of our study.  Finally, we
compare our analytical results with observed light curves of five long
gamma ray bursts and show that our model is in very good agreement with
observations using simple periodic variations of the injected velocity
profiles.  This simple method allows us to recover initial mass discharge
and energy output ejected during the burst.
\end{abstract}


\section{Introduction}

  Apparent superluminal knots observed along relativistic jets of quasars
and micro--quasars are generally interpreted as shock waves moving through
the jet. It was first \citet{rees66} who mathematically predicted the
apparent superluminal motions observed in extragalactic jets, due to
geometrical effects and relativistic velocities in the proper motion of
knots inside them.  The same author proposed that these observed knots
were produced by a varying velocity of the flow that moves along the jet
\citep{rees78}. Additionally, observations of blazars have been carried
through many years showing that the variability of the intensity and
polarisation are most likely generated by a transverse shock propagating
along the jet \citep[see, e.g.][]{hagen-thorn07, spada01,misra05}.
Recently, \citet{jamil08} have proposed an internal shock model for
microquasar jets in order to investigate particle acceleration and
radiation production in these astrophysical objects.  Another situation
where internal shock waves appear is the fireball model for long Gamma Ray
Bursts (GRBs). In such model, an effective mechanism for the generation
of the observed gamma rays is the existence of internal shock waves along
the associated jet which are caused by velocity variations of the outflow
\citep{rees94,piran04}.  Despite the alternative explanations to the
origin of internal shocks \citep[e.g.][]{narayan08} the fireball model
has been widely accepted and we show in this article how observations
match quite well with such picture.

  Detailed numerical models explaining the origin and the characteristic
radiation associated to internal shocks have been developed
\citep{blandford79,hughes85,marscher96}, with special attention to
the polarisation of the observed radiation as a probe to the internal
shocks accelerating the material.  In this article we explore the formation
of internal working surfaces propagating along a relativistic jet
originated from the periodic variation of the source velocity and/or mass
injection. Looking at the one--dimensional propagation of mass particles,
we present an extension of the non--relativistic model formulated by
\citet{canto00}. In that work, the formation and evolution of internal
working surfaces is modelled for the ejecta corresponding to stellar
jets \citep{raga92,raga04}. This model has the advantage of providing
analytic expressions for the kinetic power radiated by the mass particles
colliding inside the working surface. The extension of such analysis
to the extreme relativistic regime is presented here. Assuming that
the pressure gradients between the fluid particles are negligible,
and that radiation timescales are much shorter than the time it takes
to form a particular working surface, we are able to recover analytic
expressions for the speed of the particles at both shock fronts and for
the luminosity of the shocked gas. Our analytic model is compared to
our new numerical aztekas code (\url{www.aztekas.org}) which completely
solves the equations of hydrodynamics in a single dimension for the
relativistic regime.  In the light of our simulation, the limits for the
validity of our model are discussed.  A final and most important test
for our model is the comparison with gamma ray observations from long
GRBs. We present five cases in which the luminosity can be reproduced
with our analysis and report on the inferred physical conditions of the
ejecta that give way to the observed blast waves.

The article is organised as follows. In section \ref{dynamics} the
dynamics of the setup are discussed and the analytical description
of the problem is presented. In section \ref{constant-discharge} we
provide an example for the case of a constant mass discharge with an
oscillating initial velocity. In section \ref{numerics} we present the
numerical method used to solve the equations in one dimension with planar
symmetry. The comparison with the analytical model is presented through
the example of section \ref{constant-discharge}. As a test for the limits
of our analytic model we also run simulations where the pressure of the
fluid is non--vanishing. In section \ref{numerics} we compare the results
of the numerical and analytic methods with each other and discuss the
applications of our model.  Finally, in section \ref{astro-app} we use
our simple analytical model to fit the light curves of five long GRBs.

\section{Dynamics of relativistic working surfaces}
\label{dynamics}

  The formation of shocks along the structure of a relativistic
jet has been explained by several phenomena such as the presence
of inhomogeneities in the surrounding media, the deviations and
changes in the geometry of jets, and the temporal fluctuations
in the parameters of the ejection \citep[see e.g.][and references
therein]{rees94,mendoza-phd,mendoza-1-01,mendoza-1-02,jamil08}.  Here we
are concerned with the last situation. When the speed of the emitted
mass particles varies with time, a faster but later fluid parcel
eventually hits an earlier but slower ejection producing an initial
discontinuity which gives rise to a working surface, i.e.  a contact
discontinuity surface, two shock waves, and two regions of shocked
flow that must be at rest with respect to the contact, as the shocks
recede from the contact surface in its frame of reference.  The working
surface travels along the jet with an average speed \( v_\text{ws} \), as
measured in the frame of the jet source.
This picture for the formation of radiation shock surfaces is known
as the internal shock model \citep{rees94,daigne98}. Although several
extensions and particular aspects of the model have been presented in
the literature \citep[see e.g][]{spada99,spada01,misra05}, there are
no simple analytical descriptions of this phenomenon.  Here we present an
analytical approximation to the formation of an internal working surface
along a relativistic jet. We assume that the radiation timescales are
small compared to the characteristic dynamical times of the problem
\citep{spada01}. In consequence the pressure of the fluid is negligible
and the collision is described ballistically. This assumption is valid
if the flow within the jet is nearly adiabatic and non--turbulent
\citep{misra05}.

  To follow the evolution of the working surfaces, we consider a source
ejecting material in a preferred direction \( x \) with a velocity \(
v(\tau)  \) and a mass ejection rate \( \dot{m}(\tau) \), both dependent
on time \( \tau \) as measured from the jet's source.

\begin{figure}
\begin{center}
  \includegraphics[width=0.40\textwidth]{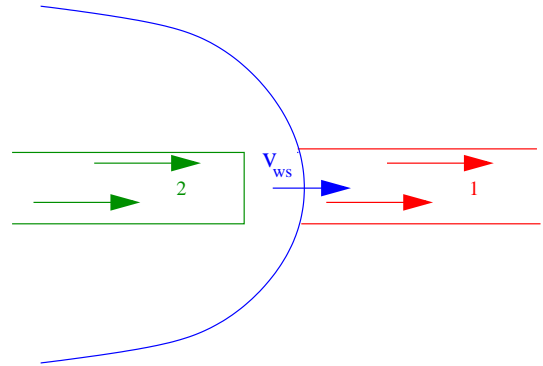}
\end{center}
  \caption{When a fast velocity flow 2  moves over a slow velocity flow
     1, a working surface (represented with a curved line)  moving
     with velocity \( v_\text{ws} \) is generated as a result of the
     interaction.  }
\label{fig01}
\end{figure}

  Once the material has been ejected from the source, we assume it will
flow in a free-stream way \citep[see e.g.][]{raga90}.  This approximation
is valid since the Mach number of the flow is large and, as mentioned
before, we emphasise that the radiation processes which cool down
the fluid occur in times scales shorter than other dynamical times
associated the to problem studied in this article \citep[cf.][]{spada01}.
The formation of a working surface is studied as the intersection of two
different parcels of material ejected at times \( \tau_1 = 0 \) and \(
\tau_2 \) with flow velocities \( v_1= v(\tau_1) \) and \( v_2 = v(\tau_2)
= v_1 + \alpha\, \tau_2 \) respectively (see Figure~\ref{fig01}), with
\( \alpha := \left( \mathrm{d} u / \mathrm{d} \tau \right)_{\tau_1} \).
If \( \alpha > 0 \), the second parcel will eventually reach the first
one. At time \( \tau_2 \), the distance between both parcels is given by
\( v_1 \left( \tau_2 - \tau_1 \right) \) and thus the time \( t_\text{m}
\), measured in the reference frame of the central engine (i.e. the
observer's frame), when both of them merge is given by

\begin{displaymath}
  \begin{split}
  t_\text{m}
      =& \frac{1}{\alpha} v_1 \gamma^2 \left( v_1 \right) \left\{ 1 -
         {v_1^2} - { \alpha v_1 \tau_2 }\right\}, \\
      =& \frac{v_1}{\alpha} \left\{ 1 - { \gamma^2( v_1 ) \,  
       \alpha \, v_1 \, \tau_2 } \right\},
  \end{split}
\end{displaymath}

\noindent where \( \gamma \left(u\right) :=  1 / \sqrt{ \left( 1
- u^2 \right) } \) represents the Lorentz factor of the flow with
velocity \( u \), and we have assumed that the speed of light \(
c = 1 \).  The working surface is then formed at a distance

\begin{displaymath}
  d_{f} = (t_\text{m} + \Delta t) v_1,
\end{displaymath}

\noindent from the central engine, where \( \Delta t =  \tau_2 - \tau_1 \).

%\emph{SERGIO: Probablemente aqui es donde podriamos describir algo
%  sobre como se forman los dos choques y como es que estan separados
%  por una discontinuidad de contacto que se mostrara mas claramente en
%  la simulacion num\'erica}

%%%%%%%%%%%%%%%%%% F I G U R E  %%%%%%%%%%%%%%%%%%%%%%%%%%%%%%%%%%%%%%
%%%%%%%%%%%%%%%%%% F I G U R E  %%%%%%%%%%%%%%%%%%%%%%%%%%%%%%%%%%%%%%

  Following the non--relativistic formalism first proposed by
\citet{canto00}, we assume that the working surface is thin and that
there are no mass losses within it (e.g. by sideways ejection of material
\citep[see][]{falle93,falle95}).   Since the flow is approximated as a
free--streaming one, its velocity \( v(x,t) \) as a function of the \( x \)
coordinate and time \( t \) is simply

\begin{equation}
  v(x,t) = v_0(\tau) = \frac{ x }{ t - \tau }.
\label{eq.1.3a}
\end{equation}

  This relation implies that the position \( x_\text{ws} \) of the
working surface from the downstream flow is given by

\begin{equation}
  x_\text{ws} = v_1 (t - \tau_1),
\label{eqn1.3}
\end{equation}

\noindent and the one corresponding to the upstream flow takes the form

\begin{equation}
  x_\text{ws} = v_2 (t - \tau_2). \label{eqn1.4}
\end{equation}

  Consistent with the assumption that the flow is free--streaming, the
velocity of the working surface is given by the velocity \( v_\text{ws}
\) of its centre of mass, which is determined by \citep{daufields}

\begin{equation}
  v_\text{ws} = \frac{ 1 }{ M_\gamma }
    \int_{\tau_1}^{\tau_2} \gamma\left(
    v(s) \right) \dot{m}(s) v(s)  \mathrm{d}s,
\label{eqn1.5}
\end{equation}

\noindent where the weighted mass \( M_\gamma \) ejected between times
\( \tau_1 \) and \( \tau_2 \) is

\begin{equation}
  M_{\gamma} = \int_{\tau_1}^{\tau_2} \gamma\left(v(s)\right) \dot{m}(s)
               \mathrm{d}s.
\label{eqn1.6}
\end{equation}

  Using formula~\eqref{eqn1.5} for the velocity \( v_\text{ws} \), it
follows that the position of the working surface is given by

\begin{equation}
  x_\text{ws} = (t - \tau_2) v_\text{ws} + \frac{ 1 }{ M_\gamma }
    \int_{ \tau_1 }^{ \tau_2 } \gamma\left( v(s) \right) \, \dot{m}(s) \,
    v(s) \, (\tau_2 - s) \, \mathrm{d}s.
\label{eqn1.7}
\end{equation}

  From equations \eqref{eqn1.3} and \eqref{eqn1.4} it follows that the
position of the working surface as a function of the times \( \tau_1 \) and
\( \tau_2 \) is also given by

\begin{equation}
  x_\text{ws} = \frac{ v_1 v_2  }{ v_2 - v_1 } \left( \tau_2 - \tau_1
    \right) .
\label{eqn1.7a}
\end{equation}

\noindent In the same manner, from the same set of equations, we
calculate the time \( t \) as a function of \( \tau_1 \) and \(
\tau_2 \), giving

\begin{equation}
  t = \frac{ \tau_2 v_2 - \tau_1 v_1 }{ v_2 - v_1 }.
\label{eqn1.7b}
\end{equation}

%%% The following statement is true since eq(1.3) = eq(1.7) given a
%%% function of \tau_1 \tau_2 and t %%%

  For a given value of the position \( x_\text{ws} \), expressions
\eqref{eqn1.3}, \eqref{eqn1.4} and \eqref{eqn1.7} establish a relation
between times \( \tau_1 \) and \( \tau_2 \).  Taking \( \tau_2 \) as
a parameter, we can construct the position and velocity of the working
surface as a function of \( \tau_2 \) and calculate the values of relevant
quantities to the problem, such as the energy available on the moving
working surface. Such a relation is one to one as long as the ejection
speed $v(\tau)$ increases monotonically. 

  In order to calculate the amount of kinetic energy radiated away as
the working surface moves, we take into account the energy \( E_0 \)
the material had when it was ejected, which is well approximated by%
% Actually, this is quite obvious.  Sergio's not sure that this should be
% left here.
\footnote{The term $\int{\dot{\gamma}(s) m(s) }\,ds$ is subdominant in
our problem as long as the variation of the velocity does not
dramatically drop the speed to very small values with respect to the speed
of light.}

\begin{equation}
  E_{0} = \int_{ \tau_1 }^{ \tau_2 } \dot{m}(s) \, \gamma\left( v(s)
    \right) \, \mathrm{d} s,
    \label{eq_e0}
\end{equation}

\noindent and the energy \( E_\text{ws} \) of the material inside the working
surface, which is given by

\begin{equation}
  E_\text{ws} = m \gamma_\text{ws},
\end{equation}

\noindent where the Lorentz factor \( \gamma_\text{ws} \) of the
working surface material is such that \( \gamma_\text{ws}^{-2} = 1
- v_\text{ws}^2 \).

  If we assume now that the energy loss along the jet, \( E_r = E_0 -
E_\text{ws} \) is completely radiated away, then the luminosity \( L :=
\mathrm{d} E_\text{r} / \mathrm{d} t \) of the working surface is given by

\begin{equation}
  \begin{split}
  % L =&  - \frac{ \mathrm{d} E_r}{ \mathrm{d} t}  \\
  L &= \frac{ \dot{m}(\tau_2) }{ \mathrm{d} t / \mathrm{d} \tau_2 }
       \left\{ \gamma_\text{ws} + \frac{ m }{ M_\gamma }
       { \gamma_\text{ws}^3 \gamma_2 } \left( v_\text{ws}
       v( \tau_2) -  v_\text{ws}^2 \right) - \gamma_2 \right\}      \\
      & - \frac{ \dot{m}(\tau_1) }{ \mathrm{d} t / \mathrm{d} \tau_2 }
       \frac{ \mathrm{d} \tau_1 }{ \mathrm{d} \tau_2} \left\{
  \gamma_\text{ws} + 
       \frac{m}{ M_{\gamma} }  \gamma_\text{ws}^3
       \gamma_1  \left( v_\text{ws} v( \tau_1) -  v_\text{ws}^2 \right)
       - \gamma_1 \right\} ,
  \end{split}
\label{eq-lumino}
\end{equation}

\noindent where the Lorentz factors \( \gamma_{1,2}^{-2} := 1 -
v^2(\tau_{1,2}) \) and, as we did before, we keep \( \tau_2 \) as
the free parameter in the expansion.  In consequence, the luminosity
\( L \) in equation \eqref{eq-lumino} is found by writing down
the expressions for \( \tau_1 \), \( v_1
\), \( v_2 \), \( v_\text{ws} \) and the derivatives \( \mathrm{d}
\tau_1 / \mathrm{d} \tau_2 \) as well as \( \mathrm{d} t /
\mathrm{d} \tau_2 \) as functions of the free parameter \( \tau_2
\).
% In this article we are concerned with the description of the working surface formation, the observational signature of such shock involves radiation mechanisms and geometrical considerations go beyond the scopes of our study. In the next section we present an example of the formation of the shock where only one of the boundary conditions is time-dependent. 

\section{A constant discharge flow}
\label{constant-discharge}

  As an example of our analytic description, let us consider the particular
case of a constant discharge \( \dot{m} \) and calculate the luminosity
\( L \) that  is obtained through simple oscillations of the particle
emission speed. We assume that the injected velocity \( v \) has a
periodic form given by 

\begin{equation}
  v(\tau)  = v_0 + \eta^2 \sin \tau.
\label{eq20} 
\end{equation}

\noindent in which the constant \( \eta \ll 1 \).
This type of oscillatory emission speeds have been widely used for the
description of internal shocks in both the Newtonian and the relativistic
cases \citep[cf.][]{canto00,spada99}.  For this example we choose a system
of units in which \( \dot{m} = 1 \).   In addition we set the time unit
so that the oscillation frequency is \( \omega = 1 \). As a consequence
of this assumption, the luminosity \( L \) is such that its dimensions
are the same as the dimensions of \( \dot{m} \), i.e.

\begin{equation}
  [L] = [\dot{m}], 
\label{eq20.0}
\end{equation}

\noindent and is thus dimensionless.

\begin{figure}
\begin{center}
  \includegraphics[width=0.50\textwidth]{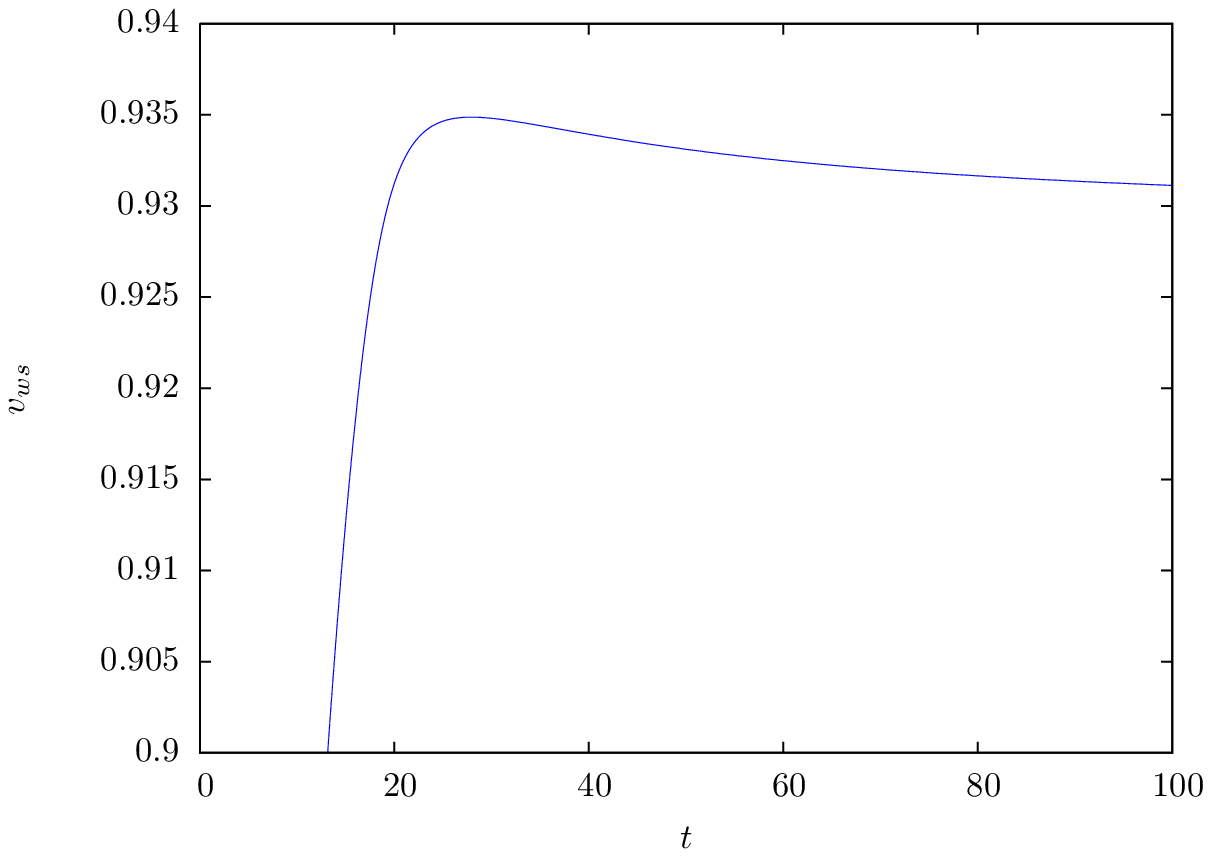} \\
  \includegraphics[width=0.50\textwidth]{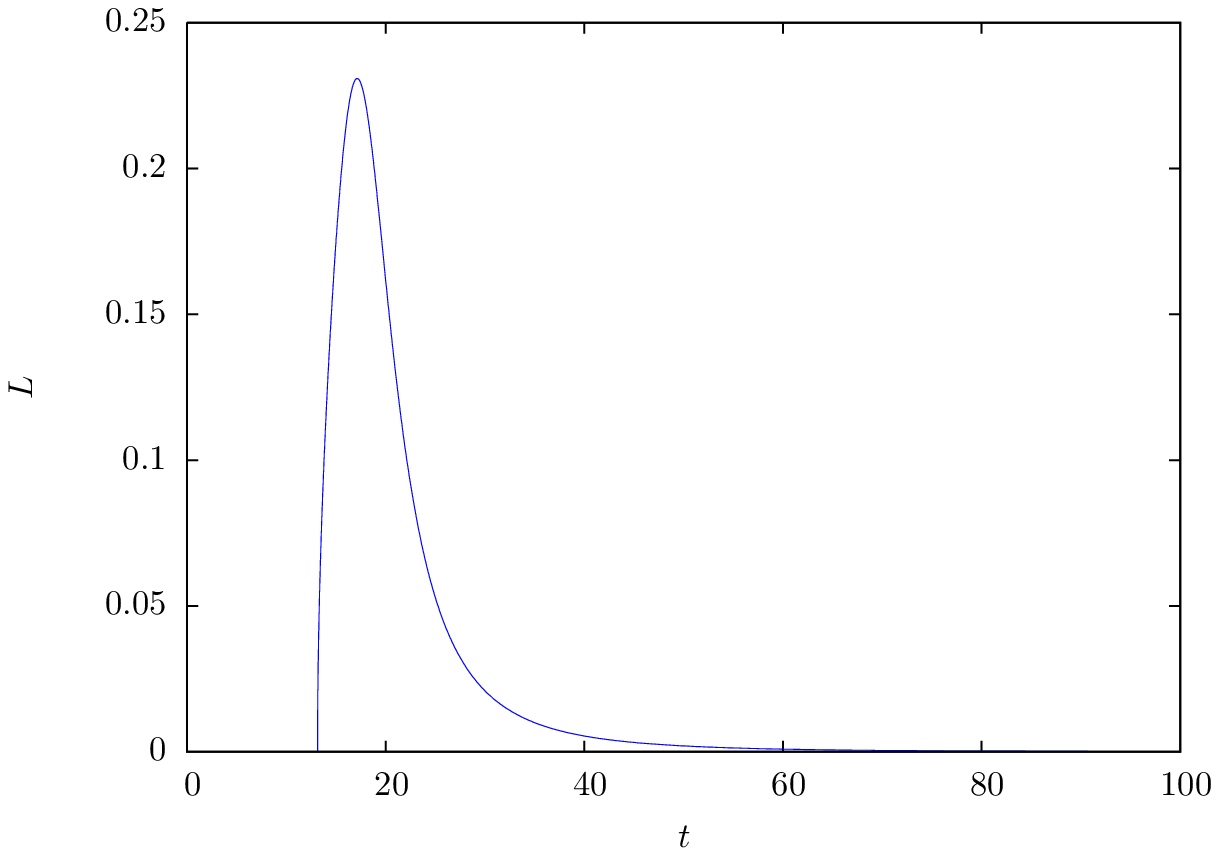}
\end{center}
  \caption{The figure shows the dimensionless velocity of the
	   working surface \( v_\text{ws} \) and its dimensionless
	   luminosity \( L \) as a function of time for a period of
	   flow oscillation with an input velocity given by
	   equation~\eqref{eq20} and a constant discharge flow, under
	   the assumption of a ballistic flow and conditions mentioned
	   in section~\ref{constant-discharge}. All these quantities are
	   measured in the rest frame of the jet source.}
\label{fig02}
\end{figure}

If one assumes that the injected flow is highly relativistic, it is
then possible to solve analytically equations
\eqref{eqn1.5}--\eqref{eq-lumino}  at \( \text{O}(\gamma^{-1})
\). However, the analytic expressions are long and cumbersome
because, as opposed to the non--relativistic case, the Lorentz factor
appears in this description ubiquitously.  We have performed numerical
integration of equations \eqref{eqn1.5} through \eqref{eq-lumino} with
a velocity profile given by equation~\eqref{eq20}, using the values \(
v_0 =0.9 \) and \( \eta^2 = 0.09 \). The results are shown in
Figure~\ref{fig02} and are very similar in shape to the ones obtained by
\citet{canto00}. 

The abrupt bump in the luminosity shows that the kinetic energy must be
radiated very effectively giving rise to emissions of high energy photons
at various wavelengths depending on the strength of the
shock. 

% Sergio says: no!  Luminosity is invariant!
%The luminosity as seen by a distant observer $L_\text{obs}$can
%be deduced performing the transformation
%\begin{equation}
%L_\text{obs} =  \frac{L}{\gamma_\text{ws}}.
%\end{equation}

The radiation mechanism and characterisation of the spectrum of
particular objects will be the subject of a subsequent article.  For a
detailed study of the radiation mechanisms of internal shocks in GRBs
see e.g. \citet{daigne98} and \citet{misra05} for the case of
radiative knots along jets in AGNs.

%%%%%%%%%%%%%%%%%% F I G U R E  %%%%%%%%%%%%%%%%%%%%%%%%%%%%%%%%%%%%%%
%%%%%%%%%%%%%%%%%% F I G U R E  %%%%%%%%%%%%%%%%%%%%%%%%%%%%%%%%%%%%%%

\section{1--D numerical solution}
\label{numerics}

  In order to compare our analytical calculations and the results
shown in Figure~\ref{fig02}, we test the validity of our model and
its approximations through a 1D Relativistic Hydrodynamic (RHD) code
developed by us for general purposes in astrophysical situations. The
code uses a finite difference method to solve the continuity, energy
and momentum equations given by \citep[see e.g.][]{blandford76,hidalgo05}

\begin{gather}
  \frac{ \partial \gamma n }{ \partial t } + \frac{ 1 }{ r^k } 
  \frac{ \partial \gamma n v }{ \partial r } = 0, 
  \label{eq002}\\
  \frac{ \partial }{ \partial t } \left( \frac{ e +v^{2} p }{
    1 - v^{2} } \right) + \frac{ 1 }{ r^k } \frac{ \partial }{ \partial r }
  \left[ r^{k} v \frac{ p +  e }{ 1 - v^{2} } \right ] = 0,
  \label{eq003}\\
  \frac{ \partial }{ \partial t }\left( v \frac{p + e}{1 -
    v^2}\right)  + \frac{1}{r^k} \frac{ \partial }{ \partial r }
  \left[ r^{k} v^2  \frac{ p +  e}{1 - v^2 }\right ] + \frac{
    \partial p }{ \partial r }  = 0. 
  \label{eq004}
\end{gather}

\noindent respectively, with \(k=0,1,2\) for planar, cylindrical and
spherical geometry. Here $n$ represents the particle number density, 
$e = n m + \epsilon $ is the energy per unit volume of the fluid, \( m := 1
\) is the average mass per particle and \( \epsilon \) the internal energy
per unit volume.  Using a Bondi--Wheeler equation of state of the form

% Given boundary and initial conditions for the
%the particle number density \( n \), the pressure \( p \) and the velocity
%\( v \) with a Bondi--Wheeler equation of state of the form 

\begin{equation}
   p = \left( \kappa - 1 \right) \epsilon,
\label{eq005}
\end{equation}

%\noindent in which \( e \) represents the energy per unit volume 
%\citep[cf.][]{daufm}, the solutions to any 1-D RHD problem can be
%found.

\noindent where \( \kappa = 4/3 \) for an  ultrarelativistic gas,
we complete a system of equations for the functions 
$n$, $p$ and $v$. The system can then be solved with suitable 
boundary and initial conditions for these quantities.

  The numerical method is based on a finite difference scheme, with
a fixed mesh and a fixed time step.  At any time, the solutions
to equations~\eqref{eq002}-\eqref{eq004} are obtained using a
\citet{maccormack} integration method.  We have tested our code with
traditional well known tests, such as the propagation of relativistic
blast waves and shock tube problems described by \citet{marti}.  
%
%\emph{Sergio: Quizas para no alardear mucho sobre el c\'odigo
%  deber\'{\i}amos de quitar el siguiente parrafo. Ahi tu lo decides}
%
The code uses the GNU Scientific Library (\url{www.gnu.org/software/gsl})
for many of its mathematical computations, since this library is very
well tested. More details on the features of this code will be published
elsewhere.  This code, named the aztekas code,  is available on the
Internet (\url{www.aztekas.org}) under a GNU General Public License (GPL),
as described by the terms of the Free Software Foundation (www.gnu.org).

  In order to make a numerical comparison with the results obtained
in section~\ref{constant-discharge}, we use planar symmetry in
equations~\eqref{eq002}-\eqref{eq004}, with a null pressure model.
At position \( x=0 \) of our domain, we inject at any time
\( t \) matter with constant mass discharge $\dot{m}$.  This injected
mass is assumed to have a velocity given by equation~\eqref{eq20}. For
simplicity, we assumme that we have a single species of particles and so,
we set a constant injected particle number per unit time flow \( \dot{n}
= 1 \), which in turn implies that the particle number density \( n \)
at  the point \( x=0 \) is given by

\begin{equation}
  n(t,x=0) = \frac{ \dot{n} }{ \gamma v(t) } = \frac{ \sqrt{ 1 - 
    v^2(t)} }{ v(t) }.
\label{eq006}
\end{equation}

\begin{figure}
\begin{center}
  \includegraphics[width=0.47\textwidth]{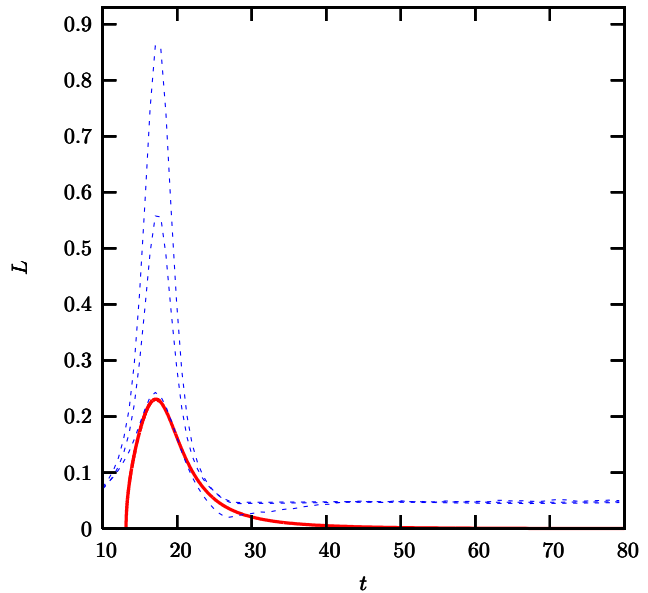}
\end{center}
  \caption{The figure shows the dimensionless luminosity \( L \) of the
	   working surface as a function of the dimensionless time \(
	   t \) for a single period (\( t = 2 \pi \)) of oscillation
	   with		  an input velocity given by equation~\eqref{eq20}
	   and a constant discharge flow.  The solid line shows the
	   analytical approximation under the assumption of zero pressure.
	   From bottom to top, the dotted curves show the numerical
	   computations made for the cases \( \zeta = 0, 0.01, 0.02 \).
	   The reason as to why the luminosity doesn't go to zero at
	   sufficiently large times (\( \gtrsim 35\)) for the numerical
	   solution is because the input velocity reaches a constant
	   value \( = 0.9 \) after a period of oscillation, whereas the
	   analytical approximation decays to a null value.}
\label{fig03}
\end{figure}

  In order to analyse a single flash of luminosity, such as the one
described in Figure~\ref{fig02},  the value of the velocity is assumed
constant after the time $t = 2\pi$, i.e. \( v(t>2\pi,x=0) = 0.9 \).
The initial conditions for the flow are chosen 
such that \( v(t=0,x) = 0.9 \), \( p(t=0,x) = 0.001 \) and \( n(t=0,x) =
\sqrt{ 1 - v^2(t=0,x) } /  v(t=0,x) \).  The shock waves obtained by
the varying velocity of the flow are captured numerically by  introducing
an artificial viscosity \citep{book75}.   Once the position of both shock
waves are known, then at each time,  the energy \( E_\text{ws} \)  of the
flow between both shock waves (the working surface) is calculated as the
sum \( \sum n \gamma \Delta x \), where the summation is done for each
numerical cell of width \( \Delta x  \) that lies between both shocks. The
energy of the input flow is calculated as in equation~\eqref{eq_e0}
and the derivative that appears in the calculation of the luminosity
\( L := \mathrm{d} \left( E_0 -E_\text{ws} \right) / \mathrm{d}t \)
is performed numerically and softened using the flux--corrector 
method presented by \citet{book75}.  Figure~\ref{fig03} presents the
numerical  result as compared with the analytic prescription described
before.  This shows that the analytical approximation is a
good way to describe the dynamics and energetics of working surfaces
formed by a varying input velocity.

  We have also performed calculations for two more cases in which the
pressure \( p \) in the hydrodynamical equations has been written as
\( \zeta p \), with \( \zeta = 0.01, \ 0.02 \).  The pressure and the
density are assumed to follow a polytropic relation of the form \( p
\propto n^{4/3} \), which is in agreement with relation~\eqref{eq005},
and \( \kappa = 4/3 \),  as described by \citet{tooper}.  The initial
and boundary conditions were not modified.  However, the input energy \(
E_0 \), given by equation~\eqref{eq_e0} is now

\begin{equation}
  E = \int_{0}^{t}{ \left[ 1 + \left( 3 + v^2 \right) \frac{ 1 }{ n }
    \frac{ \mathrm{d} p }{ \mathrm{d} t }  - \left( 3 + v^2 \right)
    \frac{ p }{ n^2 } \right] \gamma \mathrm{d} t },
\label{eq_e0-p} \end{equation}

\noindent according to equation~\eqref{eq003}.  The energy \( E_\text{ws}
\) within the working surface is calculated as the sum \( E_\text{ws}
= \sum n \gamma \Delta x + \sum p \gamma \left( 3 + v^2 \right) \Delta
x \), where the summation is taken along all cells of width \( \Delta
x \) of the domain, which lie between both shock waves.  As it can be
seen from the results presented in Figure~\ref{fig03}, the peak of the
luminosity is formed at the same time.  However, the intensity of the
pulse increases with an increasing \( \zeta \).  The case \( \zeta = 1
\), which corresponds to a full ultrarelativistic flow has not been drawn
in Figure~\ref{fig03} since its luminosity peak has a much greater value.

%%%%%%%%%%%%%%%%%% F I G U R E  %%%%%%%%%%%%%%%%%%%%%%%%%%%%%%%%%%%%%%
%%%%%%%%%%%%%%%%%% F I G U R E  %%%%%%%%%%%%%%%%%%%%%%%%%%%%%%%%%%%%%%

\section{Astrophysical applications}
\label{astro-app}

  The model developed in the previous sections can be applied to shocks
within jets emerging from AGNs, \(\mu\)-QSRs and long GRBs.  In order
to see how this simple prescription can account for some of the light
curves of these objects, we select five long gamma-ray bursts and show
that our luminosity function fits quite well their observed light curves.

  Our sample consist of five GRBs:  GRB051111, GRB060206,
GRB060904B,GRB070318 and 0GRB080413B with known redshift, observed by
the BAT instrument on board the SWIFT \citep{gehrels} satellite (taken
from the public database at \url{ftp://legacy.gsfc.nasa.gov/swift}),
in the energy range from \( 15 \) - \( 150 \) KeV.  In order to obtain
the Flux \( F \), the spectra taken at different sections of the light
curve for each individual event were adjusted as  a sample power law,
with a normalisation \( N \), and a photon index \( \alpha \) \citep[for a
more detailed explanation of spectral analysis see][and references 
therein]{firmani2008}.

  In order to fit our model to the observations, we have made use of our
null pressure analytical model with the same assumptions as the ones used
in section~\ref{constant-discharge}, but with \( v_0 = 0.99 \) and \(
\eta^2 = 0.009 \), and so the Lorentz factor of the injected flow varies
from \( \sim 50 \) to \( \sim 500 \) in an oscillating sinusoidal way.
To obtain the Flux \( F \) from the analytical approximation, we divide
the analytical Luminosity \( L \) by \( 4 \pi D^2 \), where \( D \) is the
luminosity distance, with cosmological parameters given by \( H_0 = 71 \,
\text{km} \, \text{s}^{-1} \text{Mpc}^{-1} \), \( \Omega_\text{matter}
= 0.27 \) and \( \Omega_\text{vacuum} = 0.73 \).  Note that in order to
get dimensional quantities, the dimensionless luminosity \( L \) has to be
multiplied by \( \dot{m} c^2 \), with the unknown quantity \( \dot{m} \).
In other words, we will use the fact that that \( L_\text{obs} \propto
L \), and so \( F_\text{obs} \propto F \).    In the same manner, the
dimensionless time \( t \) and the observer time \( t_\text{obs} \)
must be proportional to each other, i.e.  \( t_\text{obs} \propto t \).
The proportionality factors are obtained by a linear regression analysis
applied to both \( F \)'s and \( t \)'s separately. The results of these
fits are shown in Figure~\ref{figgrb}. Note that the observed luminosity
\( L_\text{obs} \) represents an upper limit for the luminosity, since we
have assummed that the efficiency factor \( \varepsilon \) of converting
injected kinetic energy to radiation has been taken as one.  In reality \(
\varepsilon \lesssim 1 \) and according to the results of \citet{stern08}
is close to one for Lorentz factors greater than \( 40 \).

  Complicated GRB light curve profiles may have to be adjusted by a
mixture of a sum of sinusoidal variations of not only the velocity
as shown in equation~\eqref{eq20.0}, but also as a sum of periodic
sinusoidal variations of the mass discharge \( \dot{m} \), something
outside the scope of this article.

\section{Conclusion}
\label{discussion}

  We have constructed a full relativistic solution applied to the problem
of a jet with varying periodic injection velocities and/or mass discharge.
This was done by assuming that the working surfaces formed along the jet
are ballistic.  Under these circumstances we were able to obtain a full
analytic description of their behaviour and with this, the luminosity
along the jet was calculated.  We have also made numerical comparisons
of our analytic approximations using a 1D RHD code, the current status
of the aztekas code.

  We prooved that the analytic approximations are in very good
agreement with a full numerical solution under the assumption of null
pressure gradients along the flow.  Using the numerical code, the initial
hydrodynamical quantities such as the particle discharge \( \dot{n} \),
the particle number density \( n \), the velocity \( v \) and the pressure
\( p \) can be chosen to be intricated periodic functions of time.

  We have shown how to fit astronomical observations to five long
GRBs using a simple regression analysis technique, assuming a constant
discharge flow, a simple sinusoidal variation of the velocity of the
injected flow, and a ballistic flow approximation to the problem.
It remains to test the analysis and match observations with more
complicated shapes of luminosity curves. These could be produced by
more complex flows like the sum of periodic sinusoidal functions for
the injected velocity and mass discharge. Such possibilities are left
for future research.

\section{Acknowledgements}
\label{acknowledgements}

  We thank the anonymous referee for his valuable comments, particularly
on pointing out the relevance of comparing our model with observed light
curves associated to GRBs.  We thank Jorge Cant\'o and Alex Raga for
discussions about their non--relativistic work.  We also appreciate
the comments made by William H. Lee, Xavier Hernandez and Enrico
Ramirez--Ruiz about the possible relativistic solutions to the problem.
JCH acknowledges financial support granted by CONACyT~(179026). SM \&
DO gratefully acknowledge support from DGAPA (IN119203-3) at Universidad
Nacional Aut\'onoma de M\'exico (UNAM).  SM acknowledges financial support
granted by CONACyT~(26344).  We thank the School of Mathematical Sciences,
QMUL, for their hospitality during the time of completion of this paper.

%%%%%%%%%%%%%%%%
% BIBLIOGRAPHY %
%%%%%%%%%%%%%%%%
\bibliographystyle{mn2e}
\bibliography{shock}

\begin{onecolumn}

%%%%%%%%%%%%%%%%%% F I G U R E  %%%%%%%%%%%%%%%%%%%%%%%%%%%%%%%%%%%%%%
\begin{figure}
\begin{center}
  \includegraphics[width=0.47\textwidth]{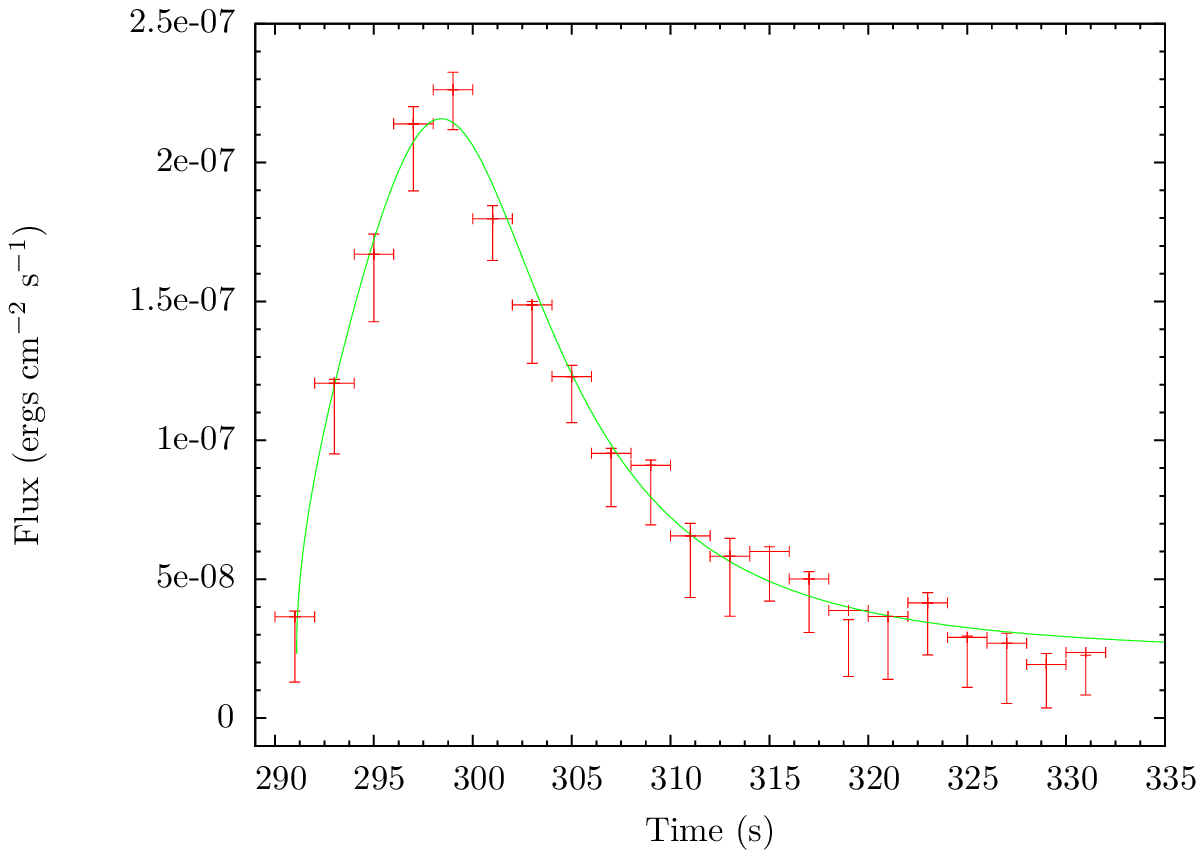}
  \includegraphics[width=0.47\textwidth]{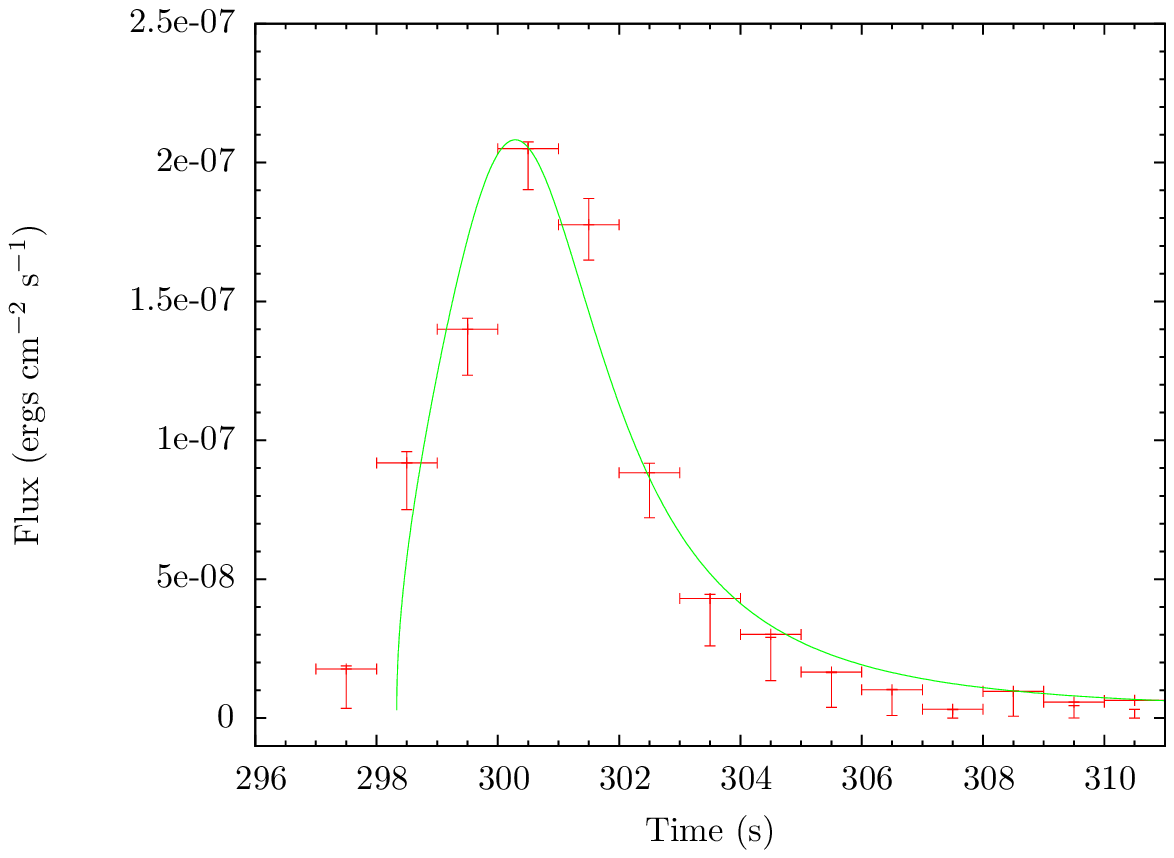} \\
  \includegraphics[width=0.47\textwidth]{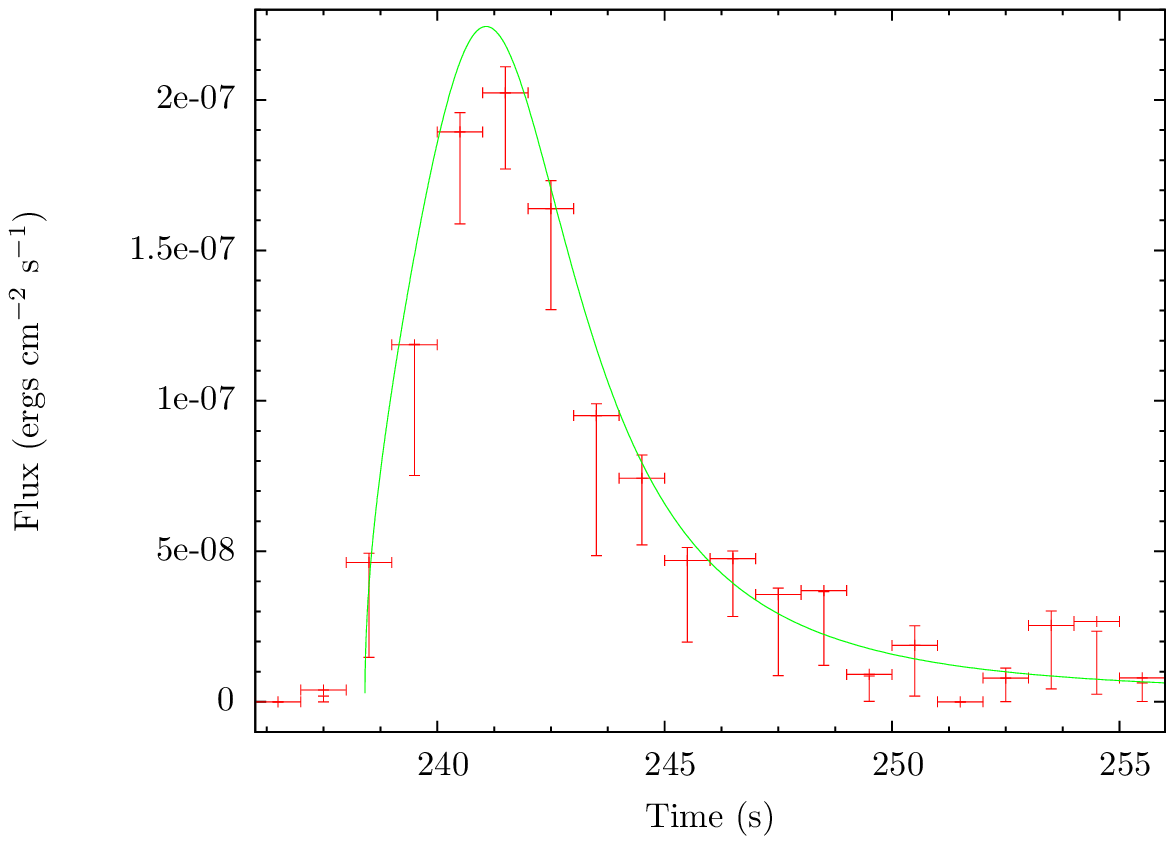}\\
  \includegraphics[width=0.47\textwidth]{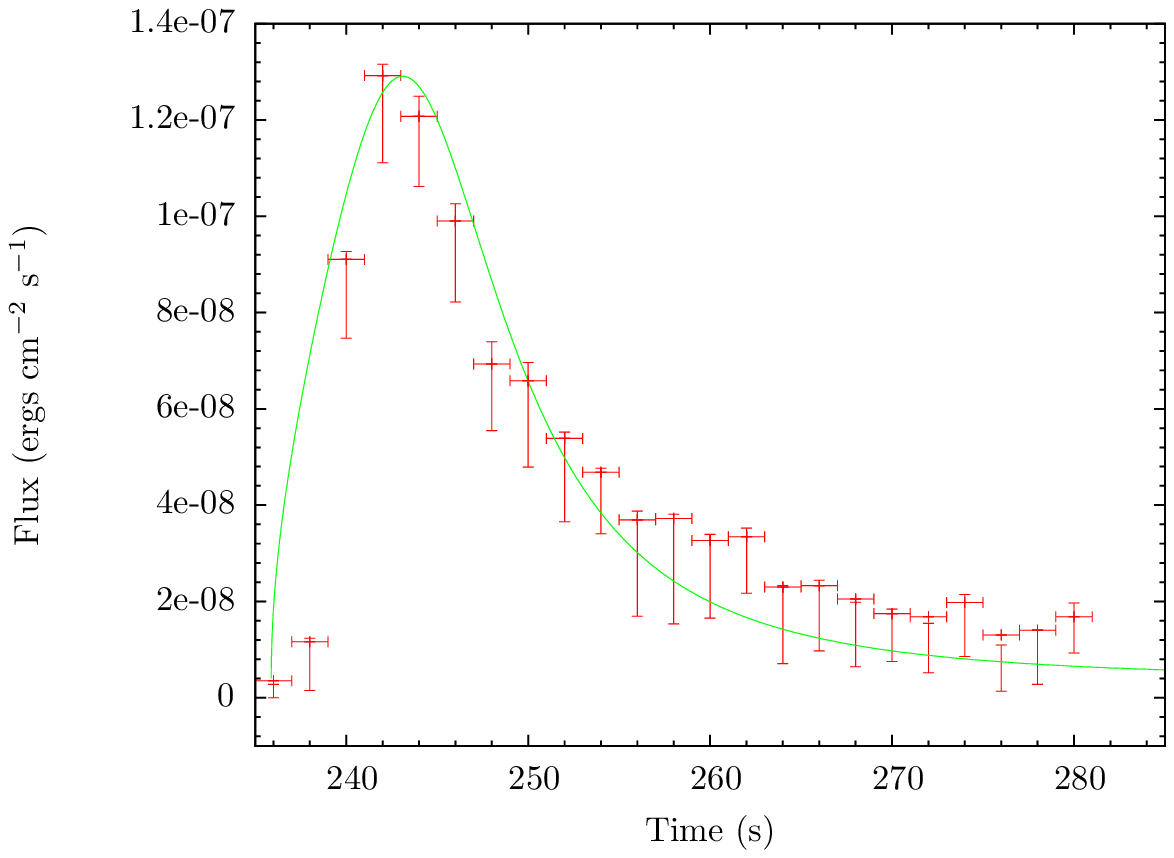}
  \includegraphics[width=0.47\textwidth]{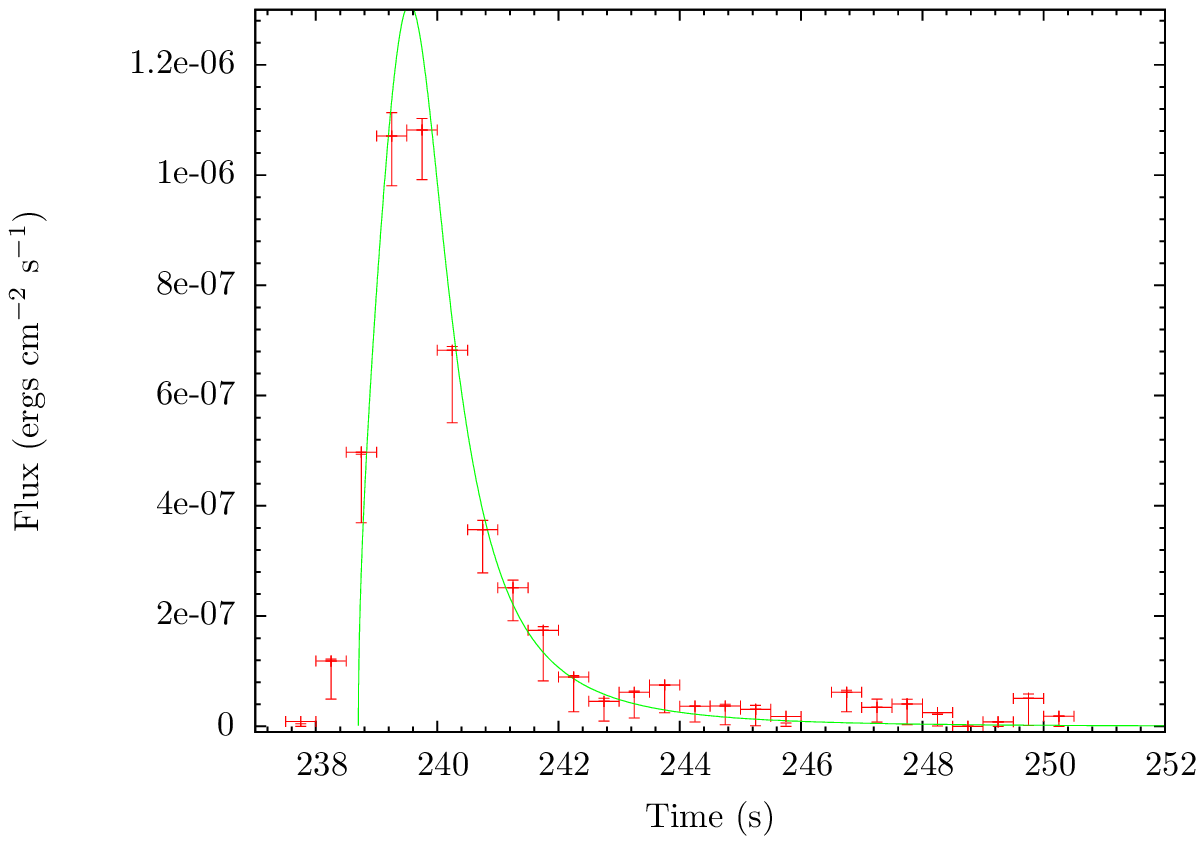}
\end{center}
  \caption{The figure shows the burst light curves 
	   (represented by dots with error bars at a \( 1\sigma \)
	   level) of five long GRBs, from top to bottom, left to
	   right, GRB051111, GRB060206, GRB060904B, GRB070318, and
	   GRB080413B observed with the BAT instrument onboard of the
	   SWIFT satellite.  The continuous curves on each graph are
	   the best adjustments using the analytical model built in
	   this article using a single period of oscillation for the
	   variation of the velocity having a sinusoidal form given by
	   equation \eqref{eq20} (see section \ref{astro-app} for more
	   details). The fits give values of the isotropic luminosity
	   given by \( \left( 12.73,\ 141.8,\ 2.042,\  1.771,\ 367.6
	   \right) \times 10^{52} \rm{ergs \ s^{-1} } \) for each burst
	   respectively, which divided by the velocity of light squared
	   imply mass injection discharges \( \dot{m} \sim \left( 10^{-1}
	   \text{ -- } 10^{-2} \right) \text{M}_\odot \, \text{s}^{-1}
	   \). }
\label{figgrb}
\end{figure}
%%%%%%%%%%%%%%%%%% F I G U R E  %%%%%%%%%%%%%%%%%%%%%%%%%%%%%%%%%%%%%%

\end{onecolumn}

% \label{lastpage}

\end{document}